# A novel coupled RPL/OSL system to understand the dynamics of the metastable states

M. Jain*, R. Kumar, M. Kook

*Department of Physics, Technical University of Denmark, DTU Risø Campus, Roskilde, Denmark*

*Corresponding author; Email: maja@dtu.dk

**Highlights**

• A unique luminescence-based physical system to track the concentration of trapped electrons and holes after thermal or optical perturbation.

• Luminescence production can suffer from a strong recombination bottleneck.

• Electron trapping probability is a function of the electron-hole distance.

• Thermal depletion of infrared stimulated luminescence reflects the availability of holes in feldspar.

**Keywords**

# Abstract

Metastable states form by charge (electron and hole) capture in defects in a solid. They play an important role in dosimetry, information storage, and many medical and industrial applications of photonics. Despite many decades of research, the exact mechanisms resulting in luminescence signals such as optically/thermally stimulated luminescence (OSL or TL) or long persistent luminescence through charge transfer across the metastable states remain poorly understood. Our lack of understanding owes to the fact that such luminescence signals arise from a convolution of several steps such as charge (de)trapping, transport and recombination, which are not possible to track individually.

Here we present a novel coupled RPL(radio-photoluminescence)/OSL system based on an electron trap in a ubiquitous, natural, geophotonic mineral called feldspar (aluminosilicate). RPL/OSL allows understanding the dynamics of the trapped electrons and trapped holes individually. We elucidate for the first time trap distribution, thermal eviction, and radiation-induced growth of trapped electron and holes.

The new methods and insights provided here are crucial for next generation model-based applications of luminescence dating in Earth and environmental sciences, e.g. thermochronometry and photochronometry.

# 1. Introduction

Metastable states in solids play an important role in dosimetry [1] and have exciting potential applications in bio-imaging, radiobiology and information storage [2-5]. These metastable states are created by the generation of free charge (electrons and holes) in a solid by exposure to ionizing radiation, followed by trapping (capture) of charge within discrete defects or defect clusters. The thermal lifetime of a metastable state (i.e. trapped electron or hole) may range from microseconds to millions of years depending on the depth of the potential well formed by electron or hole capture and the ambient temperature. Eventual detrapping by photon/phonon interactions with the traps may lead to radiative recombination of the opposite charge carriers. Depending on whether light or heat is used for detrapping (readout), the process is called optically stimulated luminescence (OSL) or thermoluminescence (TL). Prompt electron-hole (e-h) recombination at a luminescence centre, during exposure to ionising radiation, can be measured as radio-luminescence (RL). These luminescence signals may be used to estimate prior absorbed dose (J/Kg) from ionizing radiation (dosimetry), measure the burial age of sediment or rock (geochronology), map location of the emitting particles (imaging), etc.

Different metastable states are used in different applications. In $SiO_2$, for example, trapped electrons at ~2.9 eV below the conduction band edge are commonly used in OSL dating applications [6]. In long persistent phosphors, on the other hand, traps with intermediate depths (~ 0.7 eV) are required to obtain continuous emission at room temperature [7].

Despite many decades of research aimed at understanding luminescence generation from the metastable states, there remain significant gaps in our knowledge. There exists a plenitude of phenomenological and mathematical models to describe the same signal, even within the same material [1]. Largely, this ambiguity arises from the fact that luminescence signal (OSL or TL) is result of several processes: charge capture (trapping), charge release (detrapping), charge transport (localised or delocalised), and e-h recombination. Therefore, it becomes challenging to understand the exact role of these individual processes from measurement, i.e. luminescence production, resulting from the final step; often there are multiple solutions impeding an exact understanding of the physical process. This challenge can be overcome if one were able to observe independently the dynamic evolution of the trapped electron (or hole) population.

Independent measurement of trapped electrons can be made through electron paramagnetic resonance (EPR), but this technique applies only to unpaired electrons, and it is often ambiguous to relate EPR signals to the traps that participate in OSL or TL. One can also probe metastable states by optically induced intra-defect transitions (excitation → radiative relaxation) using radio-photoluminescence (RPL); the prefix radio is used to indicate that the probed states are created by ionising radiation, to distinguish it from the ordinary photoluminescence (PL). Until recently, the RPL mechanism has only been available in hole trapping states, in materials such as Ag doped crystals or glasses [8-10], and C, Mg doped $Al_2O_3$ [3]. However, in these materials, there is no direct link between OSL and RPL making the technique inappropriate for providing a holistic picture of (de)trapping and charge transport/recombination.

For investigating the physics of OSL and TL and other luminescence phenomenon based on charge transfer (e.g., long persistent luminescence), it is necessary to probe the charge-detrapping and the e-h recombination phenomena independently. A coupled RPL/OSL system based on electron traps offers such a possibility if RPL can be measured non-destructively to monitor the concentration of trapped electrons before and after an OSL measurement. One example of such a coupled RPL/OSL system is present in the $Sm^{3+}$ defect in $YPO_4$: Sm, Ce [11-13]. Here the trapped electrons in the $Sm^{2+}$ metastable state (formed by ionising radiation: $Sm^{3+} + e^- \rightarrow Sm^{2+}$) may be measured non-destructively by resonant excitation, while a higher energy excitation results in electron detrapping and subsequent transfer to the $Ce^{4+}$ hole centre (also formed by ionising radiation: $Ce^{3+} + h^+ \rightarrow Ce^{4+}$). Prasad et al. [14] used this system to obtain insights into excited-state tunnelling recombination from $Sm^{2+}$ to $Ce^{4+}$. The RPL from $Sm^{2+}$, unfortunately, quenches at around 180 K, severely restricting the investigations and applications.

More recently, our group has shown the RPL mechanism in the electron trap in feldspar [15, 16]. Metastable states in feldspar are widely used in luminescence dating and retrospective dosimetry [17-20]. *The RPL signal in feldspar, termed as infrared photoluminescence (IRPL), derives from radiative relaxation of the excited state of the main dosimetric trap (principal trap)* with a lifetime of about 30 μs at room temperature [15]. Based on low temperature spectroscopic measurements, Prasad et al. [15] concluded that both the IRPL and the OSL obtained using near infrared (NIR) excitation arise from the principal trap. *Henceforth, following the convention in luminescence dating community, the OSL induced from e-h recombination using the NIR* *excitation is referred to as infrared stimulated luminescence (IRSL)* [18]. Simplified IRPL and IRSL mechanism in feldspar are shown in Figure 1a,b; here electron detrapping followed by e-h recombination via the band tail states leads to IRSL [21-23]. Whereas, electron retrapping/relaxation to/of the excited state leads to IRPL generation. It is important to note that the probability of the excited-to-ground state relaxation is orders of magnitude higher than the recombination from the excited state [15]. Thus, at low excitation power densities (few mW.cm$^{-2}$), even at room temperature, there is negligible loss of electrons during the IRPL measurement; this is evidenced in the steady state nature (zero slope) of the IRPL signal [24]. The excitation spectrum and the Stokes shifted IRPL emission spectrum are shown in Figure 1c. Each defect emits IRPL at the rate of thousands of photons per second dependent on the excitation rate. This provides an unprecedented sensitivity for 2D [25] and 3D mapping of defect distribution and examination of charge transfer in future.

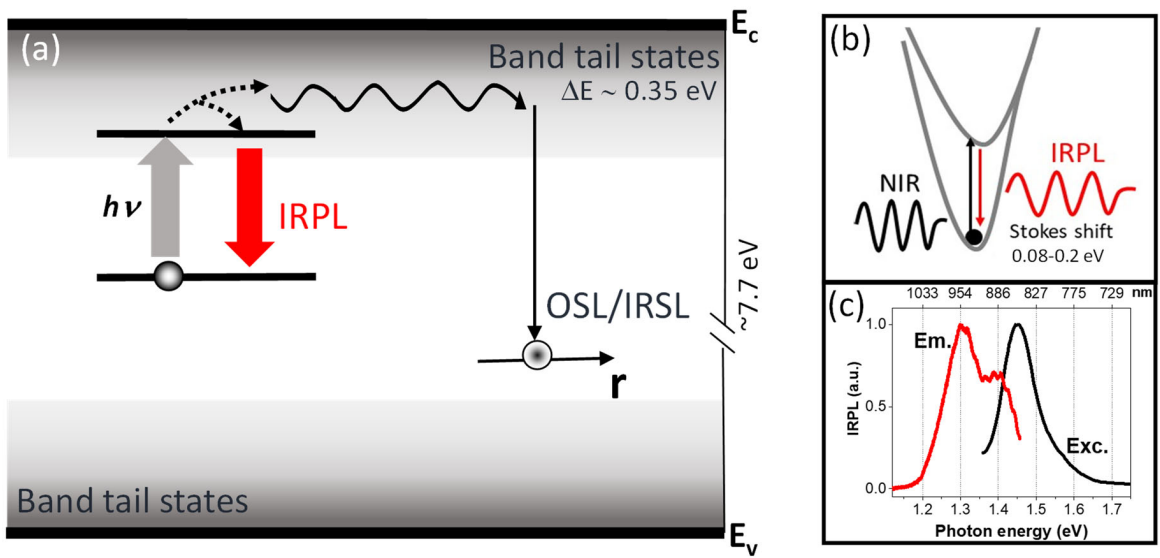

Figure 1. a) Band diagram showing the OSL (using near infrared excitation: IRSL) and the IRPL process from the principal trap; this constitutes a coupled RPL/OSL system. b) Configurational-coordinate diagram showing IRPL generation in the main dosimetric electron-trapping centre (the principal trap). The corresponding excitation (Exc.) and Stokes-shifted emission (Em.) spectra are shown in c). Microsoft PowerPoint 2013 (https://www.microsoft.com/) is used for a) and b). OriginPro 2018b (https://www.originlab.com/2018b) is used for plotting the data in c).

IRPL has the potential for providing a direct assessment of thermal stability, optical cross-sections, and trapping cross-sections of the electron traps. It is a unique tool to examine the behaviour of trapped electrons during different laboratory protocols. In this study, we explore the potential of the coupled RPL/OSL system using the IR excitation (i.e., IRPL/IRSL) to obtain a better understanding of charge transfer and recombination processes involving the principal trap. While the results here apply specifically to feldspar like systems (with both localised and delocalised charge transport), they give general insights into the behaviour of metastable states under an external stimulus.

*Note that for the rest of this article the terms RPL/OSL are used only in the generic sense; these are, respectively, the processes behind the IRPL and IRSL signals employed here.*

## 2. Current understanding of stimulated-luminescence emission in feldspar

Infrared stimulated luminescence (IRSL) is widely used in luminescence dating [26-30]. This method, however, suffers from an unexpected loss of signal (anomalous fading), a problem that has been addressed in the last decade using preferential sampling of a more stable signal. Discrimination between more and less stable signals can be achieved using a sequential measurement of IRSL at increasing sample temperatures [19,23]. Buylaert et al. [20] tested an approach using IRSL$_{290}$ after an IR$_{50}$ exposure (pIR$_{50}$IRSL$_{290}$, subscripts refer to stimulation temperature in °C) [29]. These authors found that pIR$_{50}$IRSL$_{290}$ gives an age that is consistent with the expected age supporting that such an approach can be successful in isolating stable trapped electron population. Similarly, a sequential measurement of IRSL signals at increasing temperatures gives rise to more and more stable signals [28]. Such data may be interpreted in terms of different defects with different trap depths, where the deeper defects are more stable. Equally, they may be interpreted to indicate the existence of localised recombination in feldspar [23, 31-32], explained as follows.

Around room temperature IR stimulation, the recombination primarily occurs between nearest e-h neighbours either by excited-state tunnelling or by limited diffusion within the band tail states. Such nearest neighbours are likely to recombine in nature due to tunnelling, and therefore prone to fading [33]. There remains, however, a finite population of distant e-h neighbours after this IR stimulation, for which the probability of recombination is much lower than the probability of retrapping or relaxation. If temperature is increased in a subsequent IRSL measurement, it becomes possible for detrapped electrons to access distant holes through increased thermal diffusion. The resulting post IR IRSL signals (abbreviated pIR$_T$IRSL$_T$, where T is the measurement temperature) are derived from recombination across distant e-h pairs and are therefore less prone to athermal fading (or ground state tunnelling recombination) in nature. Numerical models of feldspar suggest that intra-defect transition (excitation-relaxation) within the principal trap is the most dominant process during the resonant light excitation [31-32]; however, this transition is overlooked in the typical anti-Stokes measurements performed in OSL and IRSL. IRPL specifically measures this transition and can be read out non-destructively, especially at cryogenic temperatures. Since IRPL does not depend upon recombination, it can be measured even from electron traps that are remote from hole centres; therefore, IRPL must include a stable, steady-state component (i.e. one that does not suffer from anomalous fading) at any measurement temperature.

While this nearest-neighbour (or localised recombination) model discussed above successfully explains many experimental observations [32, 34-36], we do not have any information on how electrons-hole distances are actually distributed in the crystal. Furthermore, there is no direct proof that only a fraction of trapped electrons (from the same defect) are measured during IRSL at a given temperature, and the remaining electrons that do not participate in the IRSL are more stable. There is need for direct, unambiguous measurements to help differentiate between multiple electron trap (delocalised) model and the single trap-multiple distance (localised model). To gain better insights into feldspar model, the specific questions that we ask of a coupled RPL/OSL system are:

1) What fraction of the occupied principal trap participates in the IRSL process?
2) How do trapped electrons and/or trapped holes deplete by heat, and whether electron or hole depletion governs the thermal stability of the IRSL signal?
3) How do trapped electron and hole concentrations change due to exposure to ionizing radiation?
4) Does electron-trapping cross-section vary as a function of the e-h distance?

These aspects are investigated in the following sections. The experimental details are described in section 9. In brief, we measure both the dose-dependent Stokes-shifted IRPL emissions in feldspar (~880 and ~955 nm) in all the investigations [16]. Throughout the text, the IRPL (955 nm) and IRPL (880nm) emissions are denoted as **IRPL$_{955}$** and **IRPL$_{880}$**, respectively. Kumar et al. [16, 37] demonstrated that these signals do not represent the two excited states of the same defect site but instead two different sites; the respective (unknown) principal traps that give rise to these signals are referred to as the 880 or 955 nm (emission) centres.

### 3. Proportion of trapped electrons undergoing e-h recombination of IRSL

Here we try to experimentally test the nearest-neighbour hypothesis and determine how different sub-populations in the nearest-neighbour e-h distribution recombine in response to thermal or thermo-optical excitation. We do this by monitoring IRPL (trapped electron population) before and after IRSL (e-h recombination) at different temperatures. Seven different samples with palaeodose ranging from about 100 to 300 Gy were measured (Table 1). Three aliquots of each sample were measured using their 'natural' signals (i.e., signal due to dose received in nature) following Table 2. Subsequently, the protocol in Table 2 was repeated on these aliquots, after delivering the same beta dose as the palaeodose to avoid any dose dependent artefacts. A high temperature IR bleach (step 7) was carried out at the end of the cycle to reset the signal.

The depletion in the trapped electron population due to IR stimulation was calculated using the protocol in Table 2 as follows.

$$\Delta IRPL\ (T)\% = \frac{IRPL_i - pIR_T\ IRPL}{IRPL_i - IRPL_{bkg}} \times 100 \tag{1}$$

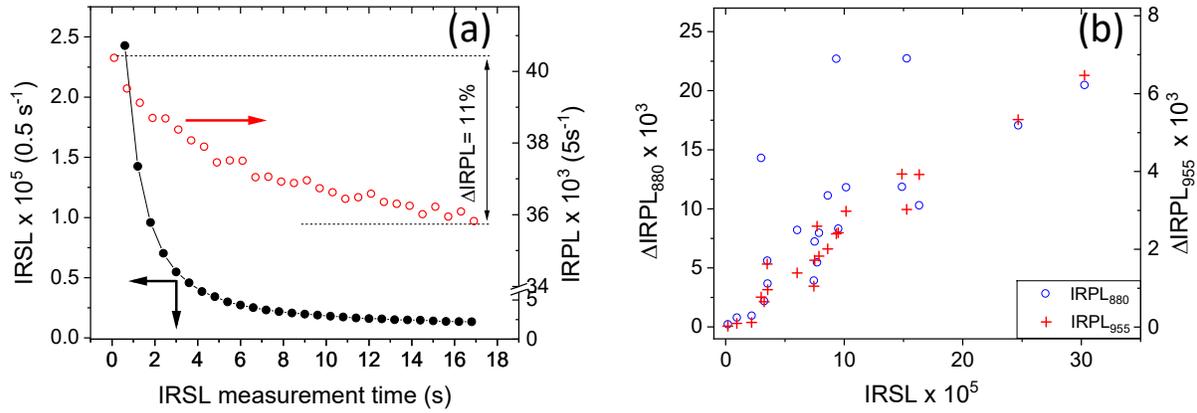

Figure 2. a) IRSL signal (left) and the corresponding changes in the intensity of the IRPL signal (right) due to the IRSL measurement. The data are produced from alternating IRSL (1s; bottom axis) and IRPL measurement cycles. The net change in the IRPL at the beginning and the end of the IRSL measurement is used to calculate the $\Delta IRPL$. b) Correlation between $\Delta IRPL_{880}$ (left) or $\Delta IRPL_{955}$ (right) and the IRSL photons emitted during the depletion of the IRPL. Three aliquots were measured from each sample. Individual data points represent one aliquot. OriginPro 2018b is used for plotting the figures: https://www.originlab.com/2018.

$\Delta IRPL$ was measured both for the $IRPL_{880}$ and $IRPL_{955}$ signals. Figure 2a shows $\Delta IRPL$ graphically for the IRSL measurement at 50 °C for the laboratory irradiated and preheated aliquots. The data in figure 2 were measured by splitting the IRSL (step 4) in 95 steps of 1 second each and monitoring IRPL after each of these steps. The IRSL signal reaches a near-constant level towards the end of the 17 s measurement, whereas IRPL (empty circles) systematically goes down to 11% of the initial value due to the IRSL measurements. The relative difference in the IRPL signals before and after the IRSL measurement is denoted as $\Delta IRPL$. $\Delta IRPL$ should reflect the population of the principal traps that participated in the IRSL (e-h recombination) process; thus, there should be a strong correlation between IRSL and $\Delta IRPL$. Figure 2b plots the relationship between $\Delta IRPL$ and net IRSL counts from the 3 aliquots each of all seven samples. We see a positive correlation between the two, however, there is a slightly greater scatter in $IRPL_{880}$ compared to $IRPL_{955}$ signal. Interestingly, the three outliers in the $IRPL_{880}$ are all from the same sample 092202. There seems to be a tendency for a slight sub-linear ΔIRPL vs. IRSL behaviour with an increase in the luminescence sensitivity; this suggests that a minor fraction of the detrapped electrons undergoes a different (i.e., not registered in our IRSL detection window) recombination route. Despite a hint of competition in e-h recombination, the majority of these data obtained from feldspar samples of different geographical origins support to a first-order approximation that ΔIRPL is proportional to IRSL.

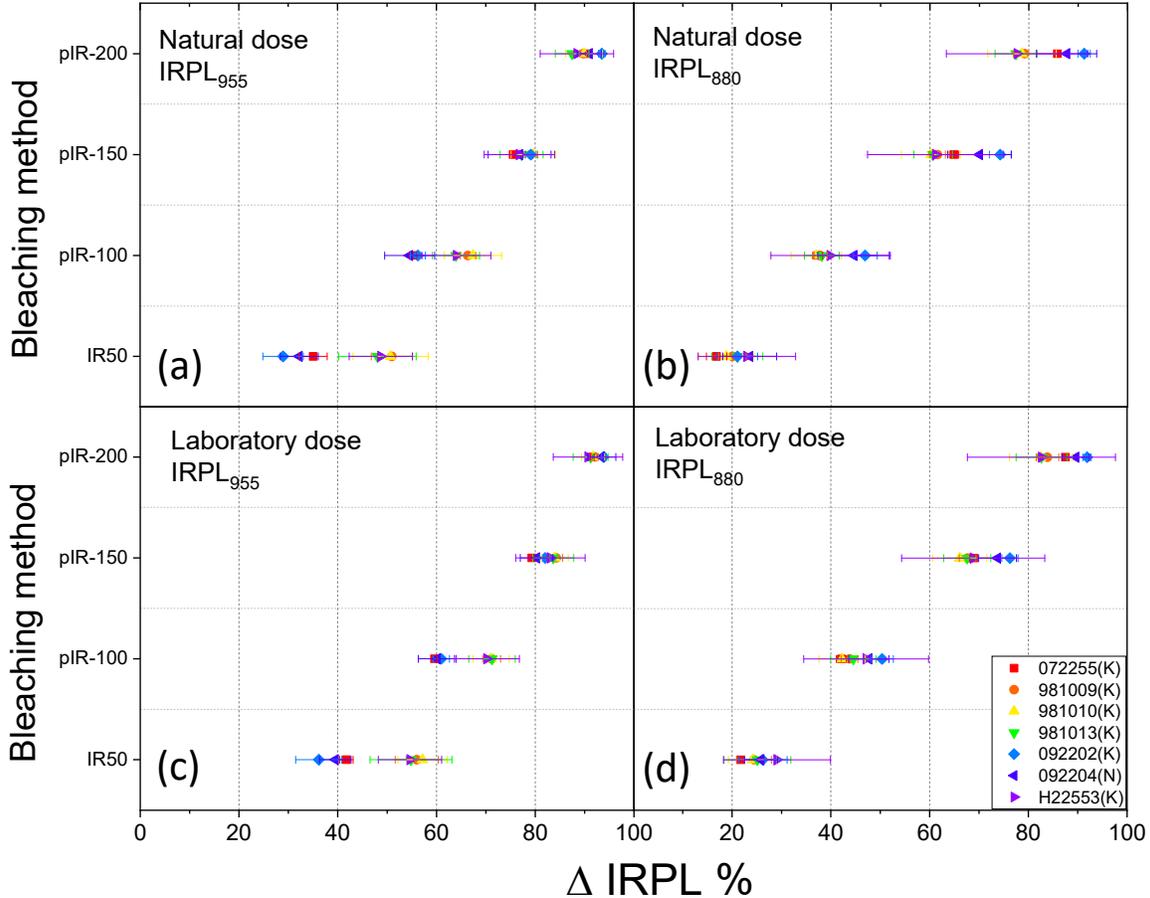

Figure 3. Δ IRPL measured after IR exposures at different temperatures (denoted as the suffices on the y-axis values). IRPL after preheat (200 °C for 60 s) but before the $IR_{50}$ exposure has been used as the baseline for the calculation. a) and b) show data for the natural dose for $\Delta IRPL_{955}$ and $\Delta IRPL_{880}$, respectively. c) and d) show data for the laboratory dose for $\Delta IRPL_{955}$ and $\Delta IRPL_{880}$, respectively. The size of the laboratory dose was kept to be the same as the natural dose for each sample. See Table 2 for details. Each data represents the average and standard deviation of three aliquots per sample. OriginPro 2018b is used for plotting the figures: https://www.originlab.com/2018.

ΔIRPL values for the IR bleach at different temperatures (step 4), estimated using the protocol in Table 2, are plotted in Figure 3. The trends between natural (Fig 3 a, b) and laboratory dose (Figure 3 c, d) are similar; however, there is a tendency for slightly larger ΔIRPL for the laboratory irradiated samples compared with the naturally irradiated samples. For $IRPL_{955}$, one can conclude that on an average about 40-50% trapped electron population participates in IRSL (step 4) at 50° C. Further, sequential raising of the IRSL temperature (sub-cycle 6) to 100, 150 or 200° C results in an average depletion of trapped electrons by about 60-70, 75-83, and 90% of the initial signal, respectively. There is a significantly large sample-to-sample spread in ΔIRPL for IR depletion at 50° C than that at the higher temperatures; this partly explains the scatter observed in Figure 2b. For $IRPL_{880}$, on an average only about 20 - 30 % of the trapped electron population participates in IRSL (step 4) at 50° C. Further sequential raising the IRSL temperature (sub-cycle 6) to 100, 150 or 200° C results in a depletion of trapped electrons by about 40-50, 65-70 and 80-90%, respectively. A significantly larger $ΔIRPL_{955}$ than $ΔIRPL_{880}$ for the $IR_{50}$ bleach explains a better correlation between IRSL and ΔIRPL for the 955 emission than the 880 nm emission; these observations show that there is a preferentially higher contribution of luminescence from the 955 centres compared over the 880 centres during the IRSL measurement.

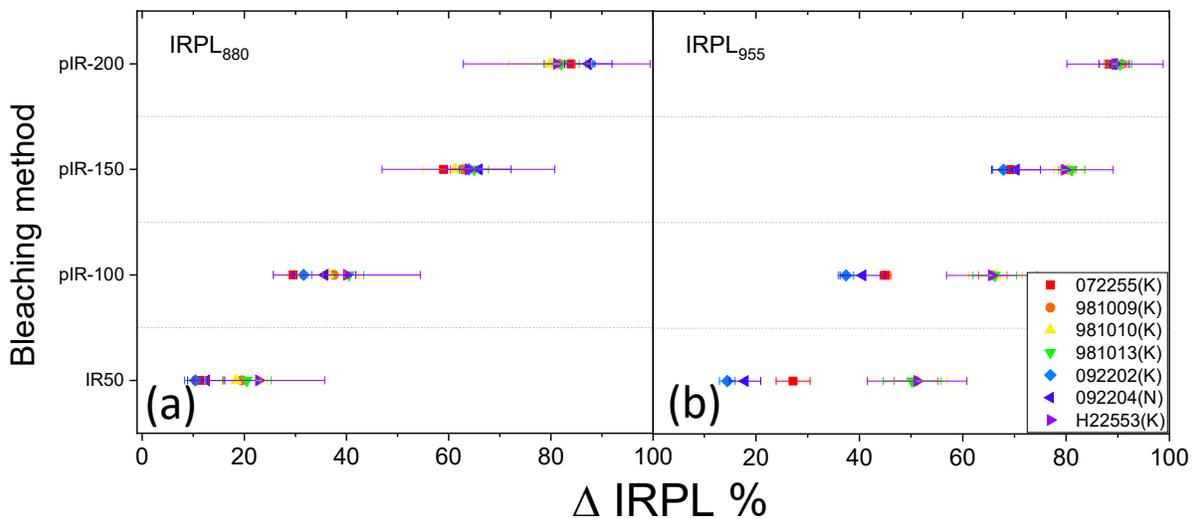

Figure 4: ΔIRPL measured after IR exposures at different temperatures (denoted as the suffices on the y-axis values). IRPL after preheat (320 °C for 60 s) but before the $IR_{50}$ exposure has been used as the baseline for the calculation. a) and b) show data for the laboratory dose for $ΔIRPL_{880}$ and $ΔIRPL_{955}$, respectively. The size of the laboratory dose was kept to be the same as the natural dose. The difference between Figure 3 (c, d) and Figure 4 is in the preheat temperature after the regeneration dose. Each data represents the average and standard deviation of three aliquots per sample. OriginPro 2018b is used for plotting the figures: https://www.originlab.com/2018.

We also measured ΔIRPL following Table 2 but with a preheat of 320 °C for 60 s, commonly used in feldspar $pIR_{50}IRSL_{290}$ dating. The results are shown in Figure 4. The $IRPL_{880}$ behaves in a similar way as in Figure 3. However, in case of $IRPL_{955}$, some samples show as low as 15% and 40% depletion in IRPL after $IR_{50}$ and $pIR_{100}$ bleaching respectively. This change likely represents a significant depletion of the unstable electron population in the $IRPL_{955}$ centre due to a 320 °C (60 s) preheat compared with 200 °C (60 s) preheat.

Our measurements of ΔIRPL are consistent with the 'nearest-neighbour' hypothesis [31] for IRSL (i.e. OSL) emission. Since IRPL is a site-selective measurement, we are always examining the same defects (880 or 955nm centres). The background level in the IRSL signal by around 17 s measurement (Figure 2a) represents a stage where e-h recombination becomes inefficient at 50 °C due to reduced access to the nearby holes (recombination bottleneck), and not due to a complete emptying of the electron trap. IRPL measurements show that for the 320 °C preheat used in feldspar dating, about 50-90% (depending on the sample) of the electron traps (880 or 955 nm centres) are still occupied when the IRSL$_{50}$ signal reaches a background level (Figure 4). *This remaining population, a function of preheat (e.g., compare Figure 3 and 4), is sampled in the pIR-IRSL methods.* The dominant process towards the end of the IR stimulation must be non-destructive excitation-relaxation (or retrapping) transitions within the still occupied principal trap (Figure 1a).

## 4. Changes in trapped electron population after preheat or high-temperature IR cleanout

It is also interesting to examine the change in the trapped electron population due to preheat (step 2) and high temperature cleanout (step 7).

For IRPL$_{955}$, the residual IRPL after a high temperature IR cleanout (IRPL$_i$-IRPL$_{bkg}$ / IRPL$_i$) ranged from about 3% to 14% in different samples, with a mean (%) ±1σ (absolute standard deviation) of 9±4 of the IRPL$_i$. The residual signal was found to be reproducible from cycle to cycle (data not shown). The change in the IRPL signal due to preheat (IRPL$_0$-IRPL$_i$ / IRPL$_0$) ranged from -14 to +5% with a mean of -2 ± 8%.

For IRPL$_{880}$, the residual IRPL after the high temperature IR bleach ranged from about 4% to 25% in different samples, with a mean of 16±8 %. The change in IRPL$_{880}$ due to preheat ranged from -20 to +1% with a mean reduction of -7 ± 9%.

The residual IRPL levels are similar to those obtained after several hours of exposure under solar simulator (data not shown) and therefore considered to represent the difficult-to-bleach trapped electron population. The change in IRPL due to preheat must arise from a combination of a) thermal depletion of the electrons in the principal trap, and b) recuperation due to electron capture in the principal trap during the decay of other shallow states. The minus sign indicates that there is a net increase in IRPL after preheat, i.e., recuperation is more dominant. The comparison of these IRPL$_{880}$ and IRPL$_{955}$ data shows that there is a greater net increase of electrons in the 880 nm traps during preheat than in the 955 nm traps. Similarly, there is a greater proportion of difficult-to-empty electrons in the 880 nm traps compared to the 955 nm traps.

## 5. Thermally-induced depletion of trapped electron and hole populations

The response of OSL to heating is commonly investigated through so called 'pulse annealing curves', where the sample is heated to different temperatures between beta or gamma irradiation and the OSL measurement [38]. Since OSL measurements involve both electrons and holes, the OSL pulse anneal curves cannot distinguish between which of the physical processes below (a-d) are responsible for the decrease in the signal. Here, $\tau_e$ is the thermal lifetime of the electrons trapped in the principal trap, and $\tau_h$ is the thermal lifetime of the holes in the recombination centre involved in OSL or IRSL. These lifetimes are related to the respective trap depths, $E_e$ and $E_h$, and the attempt to escape frequency (unit s$^{-1}$)

a. $\tau_e > \tau_h$ in a model with delocalised transport (i.e. eviction of electrons into the conduction band or holes into the valence band). The pulse anneal curve will reflect the thermal depletion of the trapped holes.

b. $\tau_e < \tau_h$ in a delocalised transport model. The pulse anneal curve will reflect the thermal depletion of the trapped electrons. This is the common conventional interpretation of the OSL or IRSL thermal depletion data.

c. Simultaneous depletion of holes and electrons in a localised transport model [31]. Here both $E_e$ and $E_h$ are significantly larger than the activation energy required to induce local e-h recombination. The pulse anneal curve will reflect the thermal activation energy for excited state tunnelling or localised recombination.

d. A localised or delocalised model with competition from shallow traps. Here both $E_e$ and $E_h$ are greater than $E_{shallow\ trap}$. Thermal eviction and subsequent recombination of electrons from the shallow traps will lead to a reduction in the trapped hole concentration; the latter will in turn lead to a decrease in the luminescence sensitivity. The pulse anneal curve will then reflect the thermal depletion of charge in the shallow trap. A complimentary scenario can be invoked for a shallow hole trap.

For a coupled RPL/OSL system one can make some predictions of the behaviour of the pulse anneal curves under these different scenarios. In a) and d), the IRPL will reduce in signal intensity at higher temperatures than for the IRSL signal. In both b) and c) the IRPL and IRSL pulse anneal curve will overlap. According to the delocalised transport model, a combination of IRPL and IRSL will provide tracking of both electrons and hole populations as a function of preheat or anneal temperature (T) as follows:

$$IRPL\ (T) \propto n_e\ (T) \tag{2}$$

$$IRSL\ (T) \propto n_e\ (T) . m_h(T) \tag{3}$$

$$\Rightarrow m_h \propto \frac{IRSL\ (T)}{IRPL\ (T)} \tag{4}$$

Here $n_e$ represents the population of the occupied principal traps, and $m$ represents the trapped hole population in the crystal, that is available for IRSL. Both are denoted as functions of preheat temperature in the above formulations.

As discussed in the previous sections 3 and 4, the IRSL process has a strong recombination bottleneck; so only a fraction of the electron population takes part in the IRSL production, e.g. at 50° C. Thus, Equations 3 and 4 are not fully justified since although IRPL originates from the entire crystal, IRSL only originates from a small sub-population that satisfy conditions for localised recombination. To tackle this problem, we also derive the thermal dependence of the IRPL lost due to the IRSL measurement, i.e. ΔIRPL (T) following equation 1. As discussed in the last section (Figure 2b), this signal should correspond to the population that participates in the IRSL process. Thus, we define two new parameters $n'_e$ and $m'_h$ to describe the behaviour of trapped electrons and holes that are active in the IRSL process:

$$n'_e(T) \propto \Delta\, IRPL\,(T) \quad (5)$$

$$m'_h(T) \propto \frac{IRSL\,(T)}{\Delta\, IRPL\,(T)} \quad (6)$$

The pulse anneal data were measured using the protocol outlined in Table 3 using three aliquots of sample 981010. Sample aliquots were first given a dose, and then heated to a certain temperature followed by immediate cool down to the room temperature (note, there is no hold at high temperature; unlike the 60 s hold for Figures 3 and 4). Subsequently, the signal was measured and the heating-cooling-measurement cycle repeated for different temperatures. The signal monitors the changes in trapped charge because of heating. The approach followed here is based on a single aliquot regenerative (SAR) dose method where any possible sensitivity change during repeated measurements is corrected for by using the response to a test dose [39,40]. From these data, we derived the following quantities as a function of anneal temperature T using the protocol outlined in Table 3:

#1. IRSL (T) / IRSL (test dose): sensitivity corrected IRSL signal.

#2. IRPL (T) / IRPL (test dose): sensitivity corrected IRPL signal. This signal measures the changes in trapped electrons in the principal trap in the entire crystal ($n_e$) as a function of preheat temperature (Equation 2).

#3. pIR$_{50}$IRPL (T) / pIR$_{50}$IRPL (test dose): sensitivity corrected pIR$_{50}$IRPL, i.e. IRPL signal measured after an IRSL measurement at 50°C. This signal measures the relatively stable principal trap population compared to those which participate in IRSL, i.e. more distant e-h neighbours.

#4. ΔIRPL (T) / ΔIRPL (test dose): this represents the thermal dependence of electrons in the principal trap ($n'_e$; equation 5), which participate in the IRSL process.

#5. IRSL (T) / IRPL (T): this represents thermal dependence of trapped holes ($m_h$; equation 4) for a delocalised model [models a), b) or d)].

#6. IRSL (T) / ΔIRPL (T): this represents thermal dependence of trapped holes ($m'_h$; equation 6) for the localised model [models c) or d)].

For all the signals (IRSL, IRPL, pIR$_{50}$IRPL) the test dose response was almost invariant as a function of the SAR cycle; nonetheless using sensitivity correction slightly improved the reproducibility. Therefore, the sensitivity corrected ratios were used as denoted above (#1, #2 and #3 and #4).

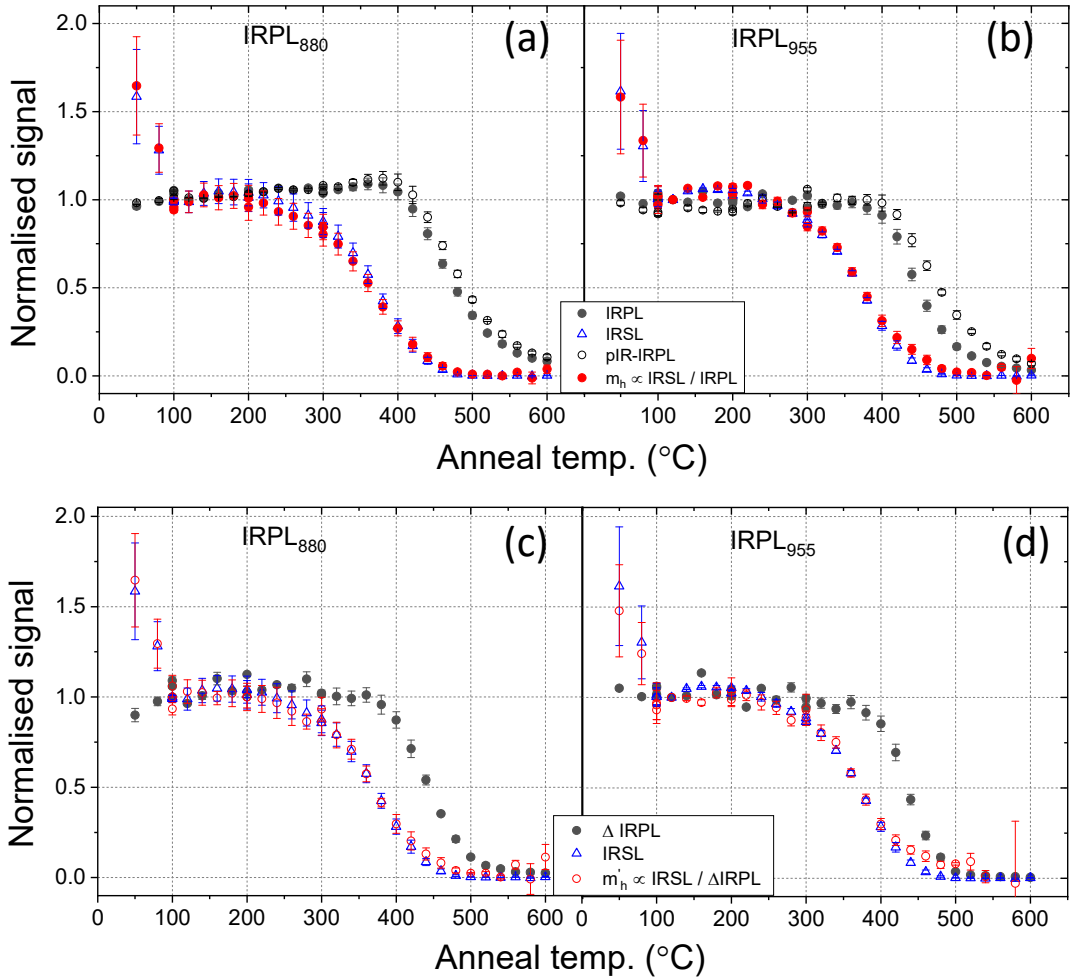

Figure 5: Pulse-anneal curves (signal vs. preheat temperature) measured for different signals using three aliquots of sample 981010. See Table 3 for details. a) Preheat dependence of the IRSL, IRPL, pIR$_{50}$IRPL signals for the 880nm emission. The behaviour of holes is calculated using equation 4. b) Preheat dependence of the IRSL, IRPL, pIR$_{50}$IRPL signals for the 955 nm emission. The behaviour of holes is calculated using equation 4. c) Preheat dependence of IRSL, ΔIRPL$_{880}$ and trapped holes (calculated using Equation 6). d) Preheat dependence of IRSL, ΔIRPL$_{955}$ and trapped holes (calculated using Equation 6). Note that the IRSL data is common for both a), b), c) and d). OriginPro 2018b is used for plotting the figures: https://www.originlab.com/2018.

These ratios (#1 to #6) are plotted in Figures 5a and c for IRPL$_{880}$ and in Figures 5b and d for IRPL$_{955}$. The data were measured on three aliquots of the sample 981010. The thermal stability of the different signals is similar for both the IRPL (880 and 955 nm) emissions. The IRSL data (signal #1 above) show a steep decrease from 50 to 100° C where it reaches a plateau between 100 and 220° C, followed by a monotonic decrease up to 450° C to a near-background value. IRPL (#2) on the other hand is relatively stable from 50 to 400° C, followed by a monotonic decrease from 400 to 600° C. Even at 600° C, ~10% of the IRPL remains. The IRSL signal has already decreased by 75% (compared to its plateau value at 100° C) at 400° C when the IRPL signal only begins to deplete. The pIR$_{50}$IRPL, i.e. the IRPL signal remaining after IR bleach (#3), is only slightly more stable than the IRPL (#1) for the 880 nm emission, while it is significantly more stable than the IRPL (#1) for the 955nm emission. This difference is not surprising since as discussed in the previous section, the change (depletion) in IRPL by IR stimulation at 50° C is much smaller for the 880 nm emission than the 955 nm emission. An increase in the stability is supported by the feldspar model, which suggests that IRSL uses the nearest e-h neighbours, which are easy to recombine by thermal stimulation through the excited state of the electron trap [31, 32].

The ratio IRSL (T) and IRPL (T) gives $m_h$ (#5). These data very closely follow the IRSL (T) depletion pattern for both the 880 and 955 emissions, suggesting that to a first-order approximation, the main cause of the decrease in IRSL is hole depletion. However, we have established in section 4 that electrons participating in the IRSL measurement (ΔIRPL) are only a fraction of the total electron population. Thus, $m'_h(T)$ (#6) is the more relevant representation of the hole population participating in the IRSL process. Interestingly, both $m_h$ (Figures 5a and b) and $m'_h$ (Figures 5c and d) show a strong overlap with the temperature dependence of IRSL. These data ($m'_h$) indicate that the IRSL pulse anneal curve is governed by the depletion of trapped holes and not the trapped electrons. Since trapped electron population is more stable than the trapped holes, we can rule out b) or c), from the four possible models discussed at the beginning of this section, as candidates governing the thermal depletion of the IRSL.

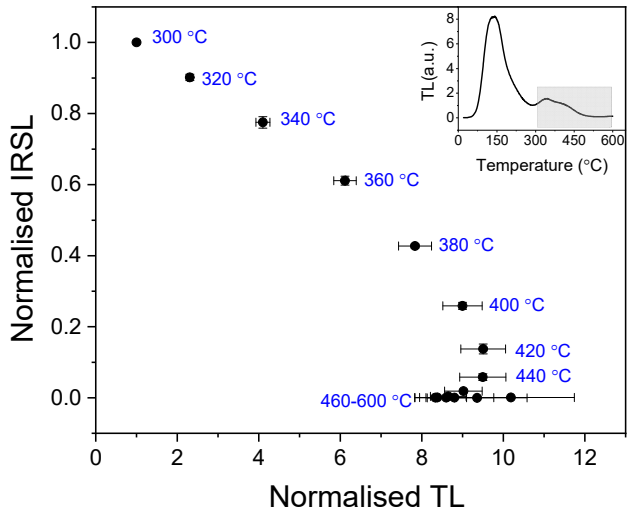

To distinguish between a) and d) as the relevant models, we examined the relationship between the TL emitted in the region 280-600° C and the subsequent IRSL counts (Figure 6). These data show a negative correlation between IRSL and TL between 300 to 400° C, i.e. the region in which the IRSL signal depletes rapidly. It is to be noted that both the IRSL and TL are measured in the same emission window

Figure 6. Correlation between integral TL intensity in the region 280° C to T° C, and the subsequent IRSL signal based on the pulse-anneal data (Table 3). The net TL signal corresponding to the IRSL was calculated from the TL curves (step 3) measured before each IRSL (step 5); this is the difference in TL counts (I) in the two integral regions ($I_{20-T°C}$ minus $I_{20-280°C}$). T is marked as the temperature against each data point. Inset shows the TL curve measured up to 600° C; the shaded area is the TL the peak used for TL - IRSL comparison. OriginPro 2018b is used for plotting the figures: https://www.originlab.com/2018.

(blue emission), thus observing the same luminescence emission centres. Since we already know that the principal trap (both IRPL and ΔIRPL) is quite stable up to a temperature of 400° C, the reduction in the IRSL must be arising from depletion of holes consumed during the TL production. These data suggest that the high temperature TL peak must be arising from an electron trap different from the principal trap and the electrons from this TL trap recombine at the same hole centres as used by the principal trap.

In conclusion, based on the novel RPL/OSL system, we are able to infer for the first time that the thermal dependence of the IRSL curve can largely be attributed to process d), i.e., depletion of holes because of competitive recombination. This is radically different interpretation of pulse anneal curves which are commonly believed to arise from the thermal erosion of the electrons in the principal trap [e.g., 41], or from localised e-h recombination [e.g., 25]. Furthermore, we establish that TL and IRSL may not arise from the same electron trap, instead both the processes use the same hole traps. Thus we infer that the commonly observed decrease in the area of the ~400 °C TL peak due to IR exposure must be due to depletion of holes. The new question that arises from our investigations is 'what mechanism results in the thermal depletion of the trapped electrons (i.e., IRPL) above ~400 °C [42] ?'. Such depletion may either be explained in the framework of delocalised or localised models and will be addressed in our future work.

## 6. Electron and hole trapping by ionizing radiation

Finally, we examine how the population of electrons in the principal trap and the holes at the IRSL recombination centres grow by exposure to beta radiation. The measurement sequence for the dose response curves (DRC) for IRPL and IRSL is outlined in Table 4.

The DRCs were measured on three aliquots of sample 981010 whose average behaviour are presented in Figure 7. Figure 7a shows the response of the sensitivity corrected IRPL signals and the IRSL signal. The dose response of the $IRPL_{880}$ is indistinguishable from that of the $IRPL_{955}$ signal. Both IRPL signals reach saturation in dose response slightly faster than the IRSL.

Figure 7b shows the dose response of the ΔIRPL signals, which is also very similar to the response of the IRPL signal; note that IRPL data from Figure 7a is also plotted for comparison. Based on the reasoning in the previous section the dose response of the holes is derived by dividing the IRSL signal by the ΔIRPL signals. These data indicate that there is about a 40% increase in the hole population from the smallest to the highest dose. The holes reach a saturation value much earlier (~0.5 kGy) than the electron trap (IRPL or ΔIRPL). Our data suggests that holes are not fully reset during a SAR cycle, and have a much lower dynamic range in dose response compared with the trapped electrons.

One question that arises is whether the dose response is similar for every component of the principal trap, or if there is a systematic variation in the DRC as we approach more and more thermally stable components. To investigate this we examined the DRC of the $IRPL_{880}$ and $IRPL_{955}$ after different preheats ranging from 360° to 480° C for 60 s. The same protocol was followed as outlined in Table 4, except for the change in thermal treatment after beta irradiation (Step 3: Preheat for 60 s at T° C; T = 360, 390, 420, 450 or 480). These data are plotted in Figures 7c and d for the IRPL$_{880}$ and IRPL$_{995}$, respectively. A linear sum of two exponential is fitted to these data

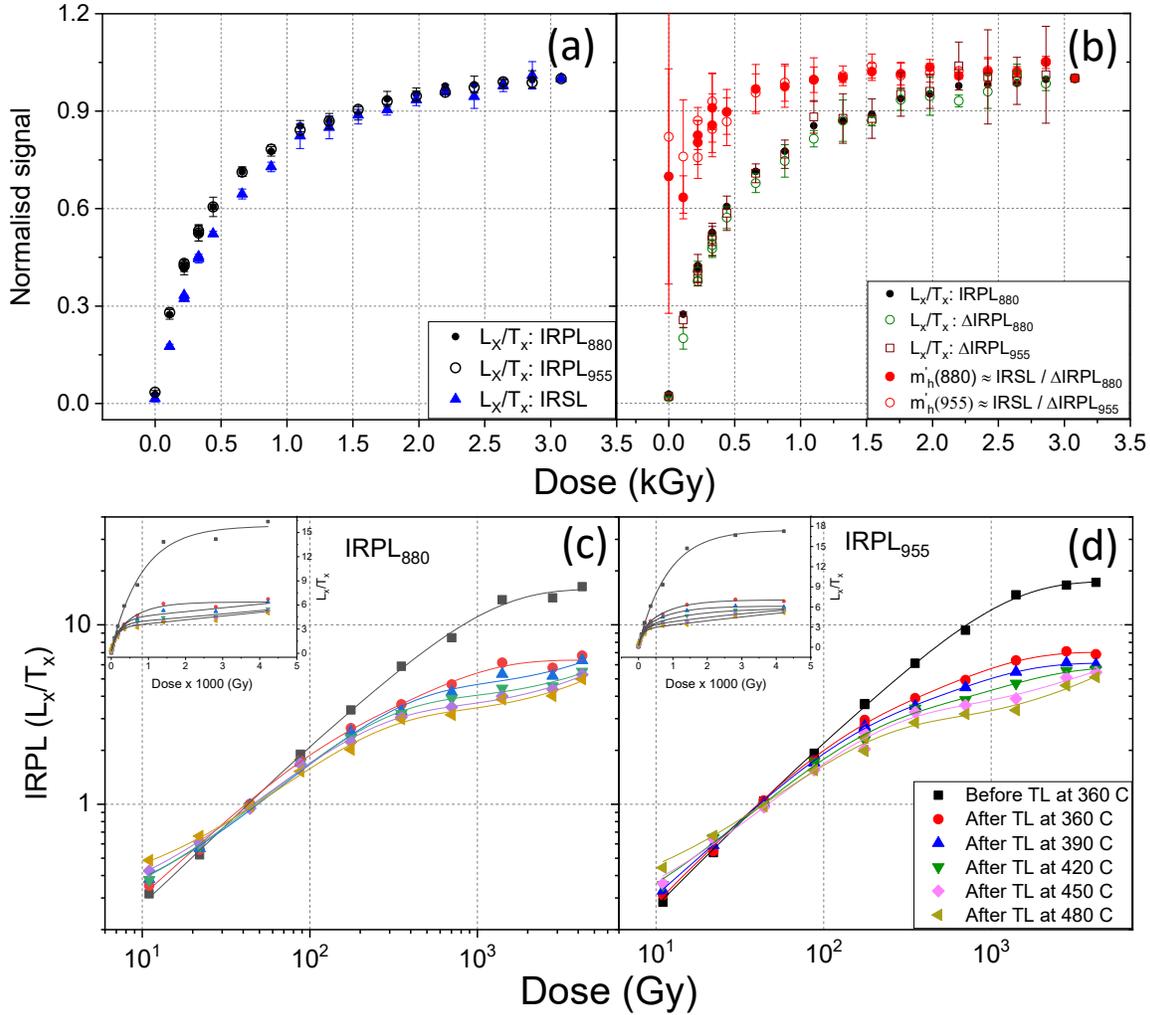

Figure 7 a) Dose response curves (signal vs. beta dose; Table 4) of the sensitivity corrected IRPL$_{880}$, IRPL$_{955,}$ and IRSL signals (steps 4-5, Table 4). b) Dose response curves of the sensitivity corrected ΔIRPL$_{880}$, ΔIRPL$_{955}$ and IRPL$_{880}$ (repeated from a) for comparison) signals. The dose dependence of holes using Equation 6 is also plotted for the 880 and 955 nm emissions. c) Dose response curve of the IRPL signals remaining after different preheats for the 880 nm emission. d) Dose response curve of the IRPL signals remaining after different preheats for the 955 nm emission. Each data point represents the mean and standard deviation from three aliquots of sample 981010. The common legend for figures c) and d) is shown in figure d). OriginPro 2018b is used for plotting the figures: https://www.originlab.com/2018.

to aid visual tracking of each signal. It is observed that there is a tendency for the DRCs to saturate earlier with dose as we access more and more thermally stable sub-populations within the principal trap. The biggest change occurs from no preheat to the preheat of 360°C (60s) scenario (Figures 7c and d). This behaviour is consistent with the response of pIR-IRSL signals, which show earlier saturation as a more stable signal is accessed [43]. In the framework of the feldspar nearest-neighbour distribution model, these data suggest that it is more difficult to trap pairs with short distances (thermally unstable) than those with large distances.

## 7. Summary and Discussion

In order to develop the future applications of the OSL/IRSL technique to understand environmental processes such as erosion and transport [44, 45], it is imperative that the luminescence kinetics under thermal or optical exposure is fully understood. Based on the coupled RPL/OSL system in feldspar, we elucidate here some long debated unknowns in the luminescence model of feldspar.

We propose a new quantity called ΔIRPL, which quantifies the fractional population of electrons in the principal trap participating in the IRSL or OSL process. The behaviour of ΔIRPL after different sample pre-treatments support the existence of recombination bottleneck in feldspar OSL discussed by [23]. These data may, however, also be interpreted in terms of a multiple trap depth model, where thermal assistance plays a role in detrapping of deeper principle traps. However, multiple-trap interpretation does not predict a systematic variation in the dose response curve with thermal stability, as is observed in our data (discussed later). Our data indicate that IRSL is preferentially derived from the principal trap emitting PL at 955 nm (IRPL$_{955}$ centre) as against that emitting at 880 nm (IRPL$_{880}$ centre). We demonstrate that ΔIRPL is a powerful tool to examine individually the behaviour of the electron and hole populations due to heating, laboratory irradiation, etc.

The measurements here give a new physical interpretation of the thermal stability of the IRSL signals. Conventional wisdom suggests that the decrease in the IRSL signal after thermal treatment (preheat or thermal bleaching) is due to eviction of electrons in the principal trap. We measure for the first time thermal response of the electrons ($n_e$, $n'_e$) and the holes participating in the IRSL process ($m_h$, $m'_h$) individually. Based on the relationship between different signals, we conclude that the thermal decay of the IRSL signal occurs mainly because the holes are used up during the 300-400 °C TL emission. Thus, the main part of the IRSL pulse-anneal curve does not represent the kinetics of the principle trap, instead it reflects the thermal decay kinetics of the TL trap. These unique first insights imply that the kinetic models [e.g., 32, 46-47] for feldspar need to be revisited in the light of these new data. This is especially important for producing robust models for thermochronometry [45, 48-49]. To arrive at a universal picture, such data need to be measured on many different types of samples.

The dose response data presented here show, for the first time, the growth of the electrons and holes separately. These data question the use of holes for dating [50], since it is apparent that holes do not exhibit a large dynamic range in dose response.

Both the IRPL and the earlier reported pIR$_T$IRSL$_T$ data show that DRC saturate earlier and earlier as we sample increasingly stable trapped charge population, e.g. by increasing the preheat (Fig. 7 c, d). In other words, it is easier to trap more stable trapped electrons, then the less stable ones. We propose a novel mechanism based on the nearest-neighbour model to explain this change in DRC of the trapped electrons as a function of preheat temperature. According to this mechanism, it becomes increasingly difficult to capture electrons in the principal trap as the e-h distance becomes smaller and smaller; this results in a higher $D_0$ for the less stable (i.e. nearer e-h) population. In other words, the capture probability is a product of both the electron capture cross-section of a trap and the distance between that electron trap and its nearest hole. The difficulty in trapping in a small nearest-neighbour volume may arise from a combination of the following:

1) The probability of creation of an e-h pair in a given volume element is a function of the pre-existing hole density. The distance from an electron to its nearest hole is called the nearest - neighbour distance. For randomly distributed holes, the nearest e-h neighbours can be described by a peak shaped probability distribution [33] as a function of the distance separating them:
$$p\,(r')dr' = 3\,(r')^2 \exp(-r')^3)dr'$$

Here $r'$ is the dimensionless distance defined as $(4\pi\rho/3)^{1/3}r$, $\rho$ is the number density of holes and $r$ is the e-h pair separation distance. Based on this model, for a random electron trapping event the relative probability of creating a nearest-neighbour at distance $r'$ is less than that for the mean distance $(\overline{r'})$, if $r' < \overline{r'}$. Thus, our IRSL dose response curves may represent the probability of e-h pair creation, rather than the probability of electron trapping alone.

2) As a free-electron approaches a principal trap at a few nm distance to a trapped hole, there is a finite probability that this electron instead of filling the electron trap may recombine with the hole. The electron and hole trapping has to then start again from the beginning to create a close distance e-h pair; thus higher dose is required to fully occupy the closely spaced, and therefore, by definition, relatively unstable traps.

The elegance of this model is that it does not invoke different principal traps to explain the variation in the dose response curves across different trapped electron populations. Instead, such a dependence comes naturally from our existing understanding of the localised recombination processes in feldspar. Thus, the apparent change in the electron capture probability for different subpopulations of the principal trap arises due to their proximity to the hole centres.

Finally, a comparison between the 880 and 955nm IRPL emissions confirms that there are two different centres (sites) comprising the principal trap. These centres show a very similar radiation-induced growth behaviour, as also observed by Kumar et al. [16]. However, they differ in their thermal stability, and ability to be bleached by IR photons (at different temperatures). These data suggest that it is likely that the principal trap consists of the same defect; however, the mean distance of this trap to the recombination sites is different for the $IRPL_{880}$ and $IRPL_{995}$ centres. Future studies involving high-resolution mapping of IRPL emitting volumes can throw light on possible dependence between microstructure or compositional variations and their effect on charge transfer to/from the principal trap. Such insights are critical for the development of exact mathematical models of luminescence phenomena involving metastable states.

## 8. Conclusions

We develop here a coupled RPL/OSL system and demonstrate its unique importance for understanding luminescence recombination pathways involving the metastable states. We suggest for the first time new quantities that can measure the behaviour of trapped electron and hole populations individually. The RPL/OSL system reveals new insights on the origins of infrared stimulated luminescence (IRSL), a commonly used OSL signal in geochronology:

1) Only a fraction of trapped electrons participates in IRSL; this observation supports a recombination bottleneck model. Thus, at any given instance, there exists a sub-population of the principal trap(s) than can be probed non-destructively using IRPL at room temperature.

2) The thermal stability of the IRSL is governed by hole depletion and not by electron depletion.

3) The net electron trapping probability in the principal trap is both a function of electron capture cross-section and its distance to the nearest hole.

These physical insights obtained from direct observation of trapped electrons have significantly enriched the feldspar luminescence model. We conceptually extend the framework of localised models to understand dose response curves. Future studies should focus on understanding the kinetics of trapping and detrapping in feldspar by applying RPL/OSL on different spatial scales. Such data will help develop robust mathematical models for next generation luminescence methods to quantify rates of Earth surface processes, e.g., thermochronometry, sediment transport, erosion, which currently work on 'black box' assumptions.

We expect that this work will inspire a search of a similar coupled PL-OSL systems in other solid-state dosimeters and enrich our understanding of charge transfer and luminescence phenomena involving the metastable states.

## 9. Materials and methods

Samples used in this study consist of K-feldspar (7) and Na-Feldspar (1) extracts from sediment samples (Table 1). These samples, 072255(K), 981009(K), 981010(K), 981013(K), 092202(K), 092204(N) and H22553(K) have been reported in Buylaert et al. [20]. Note that we have added here (K) or (N) to indicate the composition, potassium or sodium, respectively; the compositions were confirmed with X-ray fluorescence (XRF) measurements. We have deliberately chosen sediment samples, because in our experience these samples generally consist of mature minerals that have survived the physical and chemical weathering processes, and because the results of our investigations here are relevant for OSL and TL dating applications, which primarily use sediments as samples.

All measurements were performed using the photomultiplier (PMT) based IRPL attachment to the Risø TL/OSL reader. This attachment consists of an external laser light source at 1.49 eV (830 nm). Dose-dependent Stokes-shifted IRPL emissions in feldspar is measured at at ~880 and ~955 nm [16].Two different photo-multiplier tubes in combination with the emission band-pass interference filters transmitting 880/10 nm or 950/50 nm were used for detecting the IRPL emissions at ~1.41 eV (880 nm) and ~1.30 eV (955 nm), respectively [24]. The power density of the laser at the sample position was measured to be ~3 mW.cm$^{-2}$. The IRPL signals were measured in pulsed excitation mode (laser pulse width 50 µs and pulse period 100 µs; IRPL detection during 51-100 µs). Room temperature refers to the controlled laboratory temperature of 25 °C.

The OSL signals obtained using a NIR excitation (i.e. 850 nm), known as infrared stimulated luminescence (IRSL) were detected using the same PMT as the IRPL$_{880}$ but using BG39 and BG3 filters. IR light-emitting diodes (LEDs; power density ~250 mW.cm$^{-2}$ at the sample position) were used as the excitation light source. The switch over between different filters and detectors was achieved using the automated detection and stimulation head (DASH) [51].

Each sample aliquot for IRPL or IRSL measurement consisted of about 500 grains of about 150µm diameter. Both IRPL and IRSL data were analysed using the Matlab and Microsoft Excel software. OriginPro 2018b is used for model fitting and for plotting the figures.

**Table 1:** Feldspar samples investigated in this study. These samples are extracted from sediments from different geographical regions. (K) denotes K-feldspar and (N) denotes Na-feldspar. Samples and their extraction procedure is described in Buylaert et al. (2012).

| Sample code (K- or Na- Feldspar) | Site and Location | Grain size (μm) | Known $D_e$ (Gy) |
|---|---|---|---|
| 981009 (K) | Gammelmark (Denmark) | 150-250 | 279±11 |
| 981010 (K) |  | 150-250 | 298±12 |
| 981013 (K) |  | 90-250 | 274±12 |
| H22553 (K) | Sula (Russia) | 180-250 | 209±11 |
| 072255 (K) | Carregueira (Portugal) | 180-250 | 97±7 |
| 092202 (K) | Indre-et-Loire (France) | 180-250 | 158±10 |
| 092204 (Na) | Sinai peninsula (Egypt) | 180-250 | 97±3 |

**Table 2.** Measurement of depletion in IRPL due to preheat and due to IRSL at different temperatures. 'β' denotes the heating rate. 'p' denotes the holding time (pause) after reaching the desired end-temperature before switching on the light.

| Step no | Measurement | Signal |
|---|---|---|
| 0 | Prepare naturally irradiated aliquots | |
| 1a | IRPL (880 nm) for 5 s | $IRPL_0$ *(880)* |
| 1b | IRPL (955 nm) for 5 s | $IRPL_0$ (955) |
| 2 | Preheat (200° C) for 60s (β = 5° C.s$^{-1}$) | |
| 3a | IRPL (880 nm) for 5 s | **$IRPL_i$ (880)** |
| 3b | IRPL (955 nm) for 5 s | **$IRPL_i$ (955)** |
| 4 | **IRSL at T° C for 95 s (β = 5 °C.s$^{-1}$; p = 5s )** | Bleaching using IR |
| **5a** | IRPL (880 nm) for 5 s | **$pIR_T$ IRPL (880)** |
| 5b | IRPL (955 nm) for 5 s | **$pIR_T$ IRPL (955)** |
| 6 | Repeat steps 4-5 for T=50, 100, 150, 200, 250 | |
| 7 | IRSL at 290° C for 95 s (β = 5° C.s$^{-1}$; p = 5s ) | Cleanout |
| 8a | IRPL (880 nm) for 5 s | $IRPL_{bkg}$ (880) |
| 8b | IRPL (955 nm) for 5 s | $IRPL_{bkg}$ (955) |
| 9 | Laboratory irradiation | |
| | Repeat steps 1-8 | |

**Table 3.** SAR protocol for the measurement of pulse-anneal curves. IRPLs $(\lambda)$ refer to IRPL (880 nm) followed by IRPL (955nm). 'β' denotes the heating rate. 'p' denotes the holding time (pause) after reaching the desired end-temperature before switching on the light. R refers to the repeat measurement for testing the reproducibility.

| Step no | Measurement | Signal |
|---|---|---|
| 1 | Beta irradiation 110 Gy | |
| 2 | IRPLs (λ) at 20°C for 10 s | |
| 3 | TL to T° C (β = 10°C.s$^{-1}$) | |
| 4 | IRPLs (λ) at 20°C for 10 s | **L$_x$ IRPL$_\lambda$** |
| 5 | IRSL at 30° C for 100 s (β = 5°C.s$^{-1}$; p = 5s ) | **L$_x$ IRSL** |
| 6 | IRPLs (λ) at 20°C for 10 s | **L$_x$ pIR-IRPL$_\lambda$** |
| 7 | IRSL at 290° C for 100 s (β = 5°C.s$^{-1}$; p = 5s ) | Cleanout |
| 8 | IRPLs (λ) at 20°C for 10 s | IRPL$_{bkg}$ |
| | | |
| 9 | Test dose (td) of 110 Gy | |
| 10 | IRPLs (λ) at 20°C for 10 s | |
| 11 | TL to 250° C (β = 10°C.s$^{-1}$) | |
| 12-16 | repeat steps 4-8 | **T$_x$ IRPL$_\lambda$**; **T$_x$ IRSL**; **T$_x$ pIR-IRPL$_\lambda$**; IRPL$_{bkg}$ |
| | Repeat the entire cycle (steps 1-16) for T= 50, 80, 100, 120, 140, 160, 180, 200, 220, 240, 260, 280, 300, 320, 340, 360, 380, 400, 420, 440, 460, 480, 500, 520, 540, 560, 580, 600, 100R, 200R, 300R, or 100R  °C | |

**Table 4.** SAR cycle for the measurement of dose response curves. IRPLs *(λ)* refer to IRPL (880 nm) followed by IRPL (955nm). 'β' denotes the heating rate. 'p' denotes the holding time (pause) after reaching the desired end-temperature before switching on the light.

| Step no. | Measurement | Signal |
|---|---|---|
| 1 | IRSL at 290° C for 95 s (β = 5°C.s$^{-1}$; p = 5s ) | Cleanout |
| 2 | Beta irradiation (regeneration dose) | |
| 3 | Preheat (320° C) for 60 s (β = 5°C.s$^{-1}$) | |
| 4 | IRPLs (λ) at 20°C for 10 s | **L$_x$ IRPL$_λ$** |
| 5 | IRSL bleaching at 50° C for 100s | **L$_x$ IRSL** |
| 6 | IRPLs (λ) at 20°C for 10 s | **L$_x$ pIR-IRPL$_λ$** |
| 7 | IRSL at 290° C for 95 s (β = 5°C.s$^{-1}$; p = 5s ) | Cleanout |
| 8 | IRPLs (λ) at 20°C for 10 s | IRPL$_{bkg}$ |
| 9 | Test dose 220 Gy | |
| | Repeat steps 3-8 to monitor possible sensitivity changes | **T$_x$ IRPL$_λ$; T$_x$ IRSL; T$_x$ pIR-IRPL$_λ$;** IRPL$_{bkg}$ |

**References**

[1] Bøtter-Jensen, L., McKeever, S. W. & Wintle, A. G. *Optically Stimulated Luminescence Dosimetry*. (Elsevier, 2003).

[2] Liu, J. et al. Imaging and therapeutic applications of persistent luminescence nanomaterials. Advanced drug delivery reviews **138**, 193-210 (2019).

[3] Akselrod, M. & Kouwenberg, J. Fluorescent nuclear track detectors–Review of past, present and future of the technology. Radiation Measurements **117**, 35-51 (2018).

[4] Chakrabarti, K., Mathur, V. K., Thomas, L. A. & Abbundi, R. J. Charge trapping and mechanism of stimulated luminescence in CaS: Ce, Sm. Journal of applied physics **65**, 2021-2023 (1989).

[5] Meijerink, A., Schipper, W. J. & Blasse, G. Photostimulated luminescence and thermally stimulated luminescence of Y2SiO5-Ce, Sm. Journal of Physics D: Applied Physics **24**, 997 (1991).

[6] Huntley D. J., Short M. A. & Dunphy K. Deep traps in quartz and their use for optical dating. Canadian Journal of Physics **74**, 81-91 (1996).

[7] Van den Eeckhout, K., Bos, A. J., Poelman, D. & Smet, P. F. Revealing trap depth distributions in persistent phosphors. Physical Review B **87**, 045126 (2013).

[8] Miyamoto, Y. et al. RPL in alpha particle irradiated Ag+-doped phosphate glass. Radiation measurements **71**, 529-532 (2014).

[9] Yokota, R. & Imagawa, H. Radiophotoluminescent centers in silver-activated phosphate glass. Journal of the Physical Society of Japan **23**, 1038-1048 (1967).

[10] Perry, J. A. *RPL Dosimetry, Radiophotoluminescence in Health Physics*. (CRC Press, 1987)

[11] Poolton, N. R. J., Bos, A. J. J., Jones, G. O. & Dorenbos, P. Probing electron transfer processes in Y PO4: Ce, Sm by combined synchrotron–laser excitation spectroscopy. Journal of Physics: Condensed Matter **22**, 185403 (2010).

[12] Poolton, N. R. J., Bos, A. J. J. & Dorenbos, P. Luminescence emission from metastable Sm2+ defects in Y PO4: Ce, Sm. Journal of Physics: Condensed Matter **24**, 225502 (2012).

[13] Dorenbos, P., Bos, A. J. J. & Poolton, N. R. J. Electron transfer processes in double lanthanide activated YPO4. Optical Materials **33**, 1019-1023 (2011).

[14] Prasad, A. K., Kook, M. & Jain, M. Probing metastable Sm2+ and optically stimulated tunnelling emission in YPO4: Ce, Sm. Radiation Measurements **106**, 61-66 (2017).

[15] Prasad, A. K., Poolton, N. R. J., Kook, M. & Jain, M. Optical dating in a new light: A direct, non-destructive probe of trapped electrons. Scientific Reports **7**, 1–15 (2017).

[16] Kumar, R., Kook, M., Murray, A. S. & Jain, M. Towards direct measurement of electrons in metastable states in K-feldspar: Do infrared-photoluminescence and radioluminescence probe the same trap? Radiation Measurements **120**, 7–13 (2018).

[17] Huntley, D. J., Godfrey-Smith, D. I. & Thewalt, M. L. Optical dating of sediments. Nature **313**, 105-107 (1985).

[18] Hütt, G., Jaek, I. & Tchonka, J. Optical dating: K-feldspars optical response stimulation spectra. Quaternary Science Reviews **7**, 381-385 (1988).

[19] Thomsen, K. J., Murray, A. S., Jain, M. & Bøtter-Jensen, L. Laboratory fading rates of various luminescence signals from feldspar-rich sediment extracts. *Radiat. Meas.* **43**, 1474–1486 (2008).

[20] Buylaert, J. P. et al. A robust feldspar luminescence dating method for Middle and Late Pleistocene sediments. Boreas **41**, 435–451(2012).

[21] Poolton, N. R. J., Ozanyan, K. B., Wallinga, J., Murray, A. S. & Bøtter-Jensen, L. Electrons in feldspar II: A consideration of the influence of conduction band-tail states on luminescence processes. Physics and Chemistry of Minerals **29**, 217–225 (2002).

[22] Poolton, N. R. J., Kars, R. H., Wallinga, J. & Bos, A. J. J. Direct evidence for the participation of band-tails and excited-state tunnelling in the luminescence of irradiated feldspars. Journal of Physics Condensed Matter **21**, 485505 (2009).

[23] Jain, M. & Ankjærgaard, C. Towards a non-fading signal in feldspar: insight into charge transport and tunnelling from time-resolved optically stimulated luminescence. Radiation Measurements **46**, 292-309 (2011).

[24] Kook, M., Kumar, R., Murray, A. S., Thomsen, K. J. & Jain, M. Instrumentation for the non-destructive optical measurement of trapped electrons in feldspar. Radiation Measurements **120**, 247–252 (2018).

[25] Sellwood, E.L. et al. Optical bleaching front in bedrock revealed by spatially-resolved infrared photoluminescence. Scientific Reports **9**, 1-12 (2019).

[26] Balescu, S. & Lamothe, M. Comparison of TL and IRSL age estimates of feldspar coarse grains from waterlain sediments. Quaternary Science Reviews **13**, 437-444 (1994).

[27] Buylaert, J.P., Murray, A.S., Thomsen, K.J. & Jain, M. Testing the potential of an elevated temperature IRSL signal from K-feldspar. Radiation Measurements **44**, 560-565 (2009).

[28] Li, B. & Li, S. H. Luminescence dating of K-feldspar from sediments: a protocol without anomalous fading correction. Quaternary Geochronology **6**, 468-479 (2011).

[29] Thiel, C., Buylaert, J. P., Murray, A. S. & Tsukamoto, S. On the applicability of post-IR IRSL dating to Japanese loess. Geochronometria **38**, 369 (2011).

[30] Tsukamoto, S., Kondo, R., Lauer, T. & Jain, M. Pulsed IRSL: A stable and fast bleaching luminescence signal from feldspar for dating Quaternary sediments. Quaternary Geochronology **41**, 26-36 (2017).

[31] Jain, M., Guralnik, B. & Andersen, M. T. Stimulated luminescence emission from localized recombination in randomly distributed defects. Journal of physics: Condensed matter **24**, 385402 (2012).

[32] Jain, M. et al. Kinetics of infrared stimulated luminescence from feldspars. Radiation Measurements **81**, 242-250 (2015).

[33] Huntley, D. J. An explanation of the power-law decay of luminescence. Journal of Physics: Condensed Matter **18,** 1359 (2006)

[34] Angeli, V., Kitis, G., Pagonis, V. & Polymeris, G. S. Sequential two-step optical stimulation in K-feldspars: Correlation among the luminescence signals and implications for modeling parameters. Journal of Luminescence, 117425 (2020).

[35] Şahiner, E., Kitis, G., Pagonis, V., Meriç, N. & Polymeris, G. S. Tunnelling recombination in conventional, post-infrared and post-infrared multi-elevated temperature IRSL signals in microcline K-feldspar. Journal of Luminescence **188**, 514-523 (2017).

[36] Pagonis, V., Polymeris, G. & Kitis, G. On the effect of optical and isothermal treatments on luminescence signals from feldspars. Radiation Measurements **82**, 93-101 (2015).

[37] Kumar, R., Kook, M., Murray, A. S. & Jain, M. Understanding the metastable states in K-Na aluminosilicates using novel site-selective excitation-emission spectroscopy. Journal of Physics D: Applied Physics (in press). DOI. https://iopscience.iop.org/article/10.1088/1361-6463/aba788.

[38] Duller, G. A. T. A new method for the analysis of infrared stimulated luminescence data from potassium feldspars. Radiation Measurements, 23, 281-285 (1994).

[39] Murray, A. S. & Wintle, A. G. Luminescence dating of quartz using an improved single-aliquot regenerative-dose protocol. Radiation measurements **32**, 57-73 (2000).

[40] Wallinga, J., Murray, A. & Wintle, A. The single-aliquot regenerative-dose (SAR) protocol applied to coarse-grain feldspar. Radiation Measurements **32**, 529-533 (2000).

[41] Li, B. & Li, S. H. Thermal stability of infrared stimulated luminescence of sedimentary K-feldspar. Radiation Measurements **46**, 29-36 (2011).

[42] Murray, A. S., Buylaert, J. P., Thomsen, K. J. & Jain, M. The effect of preheating on the IRSL signal from feldspar. Radiation Measurements **44**, 554-559 (2009).

[43] Andersen, M. T., Jain, M. & Tidemand-Lichtenberg, P. Red-IR stimulated luminescence in K-feldspar: Single or multiple trap origin? Journal of Applied Physics **112**, 043507 (2012).

[44] Gray, H.J., Jain, M., Sawakuchi, A.O., Mahan, S.A. & Tucker, G.E. Luminescence as a sediment tracer and provenance tool. Reviews of Geophysics **57**, 987-1017 (2019).

[45] Brown, N. D. Which geomorphic processes can be informed by luminescence measurements? Geomorphology, 107296 (2020).

[46] Guralnik, B. et al. Radiation-induced growth and isothermal decay of infrared-stimulated luminescence from feldspar. Radiation Measurements **81**, 224-231 (2015).

[47] Li, B. & Li, S. H. The effect of band-tail states on the thermal stability of the infrared stimulated luminescence from K-feldspar. Journal of Luminescence **136**, 5–10 (2013).

[48] Guralnik, B. et al. OSL-thermochronometry of feldspar from the KTB borehole, Germany. Earth and Planetary Science Letters **423**, 232-243 (2015).

[49] Brown, N. D., Rhodes, E. J. & Harrison, T. M. Using thermoluminescence signals from feldspars for low-temperature thermochronology. Quaternary Geochronology **42**, 31-41 (2017).

[50] Li, B., Jacobs, Z., Roberts, R. G. & Li, S. H. Extending the age limit of luminescence dating using the dose-dependent sensitivity of MET-pIRIR signals from K-feldspar. Quaternary Geochronology **17**, 55-67 (2013).

[51] Lapp, T. et al. A new luminescence detection and stimulation head for the Risø TL/OSL reader. Radiation Measurements **81**, 178-184 (2015).

**Acknowledgments**

We thank Prof. Andrew Murray and Dr. J-P Buylaert for the feldspar samples measured here.

**Author contributions**

M.J designed this study and the experiments reported here. MJ and RK did all the measurements and data analysis. MK developed the instrumentation. MJ wrote the manuscript. All the authors reviewed and commented on the manuscript.

**Competing Interests**

The authors declare no competing interests.